\documentclass[10pt,twocolumn]{IEEEtran}
\usepackage{soul,xcolor}
\usepackage[linesnumbered,ruled,vlined]{algorithm2e}

\SetCommentSty{mycommfont}
\usepackage{algorithmicx,comment,verbatim}
\usepackage{algpseudocode}
\usepackage{amsfonts}
\usepackage{amsmath}
\usepackage{upgreek}
\usepackage{amssymb}
\usepackage[ansinew]{inputenc} 
\usepackage{dsfont}
\usepackage{mathtools}
\usepackage{graphicx}
\usepackage[skip=1pt,font=scriptsize]{subcaption}
\usepackage{import}
\usepackage{multirow}
\usepackage{cite}
\usepackage[export]{adjustbox}
\usepackage{breqn}
\usepackage{mathrsfs}
\usepackage{acronym}
\usepackage[keeplastbox]{flushend}
\usepackage{setspace}
\usepackage{stackengine}
\usepackage{steinmetz}
\usepackage[skip=2pt,font=footnotesize]{caption}

\graphicspath{{./figures/}}
\title{Adaptive 
Millimeter-Wave Communications
Exploiting Mobility and Blockage Dynamics
}

\author{Muddassar Hussain$^\dag$, Maria Scalabrin$^\ddag$, Michele Rossi$^\ddag$, and Nicol\`{o} Michelusi$^\dag$
\thanks{$^\ddag$School of Electrical and Computer Engineering, Purdue University, email: \{hussai13,michelus\}@purdue.edu}
\thanks{$^\dag$Dept.\ of Information Engineering, University of Padova, email: \{scalabri, rossi\}@dei.unipd.it}
\thanks{An extended version of this paper appears in \cite{TVT2020}.}
\thanks{This research has been funded in part by NSF under grant CNS-1642982.}
\vspace{-7mm}
}

\IEEEoverridecommandlockouts

\begin{document}
\setstcolor{red}
\setulcolor{red}
\setul{red}{2pt}

\maketitle
\begin{abstract}
Mobility may degrade the performance of next-generation vehicular networks operating at the millimeter-wave spectrum: frequent loss of alignment and blockages require repeated beam training and handover, thus incurring huge overhead. In this paper, an adaptive and joint design of beam training, data transmission and handover is proposed, that exploits the mobility process of mobile users and the dynamics of blockages to optimally trade-off throughput and power consumption. At each time slot, the serving base station decides to perform either beam training, data communication, or handover when blockage is detected. The problem is cast as a  partially observable Markov decision process, and solved via an  approximate dynamic programming algorithm based on PERSEUS~\cite{DBLP:journals/corr/abs-1109-2145}. Numerical results show that the PERSEUS-based policy performs near-optimally, and achieves a 55\% gain in spectral efficiency compared to a baseline scheme with periodic beam training. Inspired by its structure, an adaptive heuristic policy is proposed with low computational complexity and small performance degradation.
\end{abstract}
%
%

\vspace{-5mm}
\section{Introduction}
\label{sec:Introduction}
Millimeter-wave (mm-wave) is a leading candidate to support the high capacity demands of future vehicular communications \cite{choi2016millimeter}. However, communication at these frequencies relies on highly directional transmissions and it is highly susceptible to blockages and mis-alignment.  
These features are exacerbated in highly mobile environments, resulting in degraded system performance. 
 To compensate for these effects, 
the key question addressed in this paper is the following: \emph{How can we leverage the information on the system dynamics (mobility of users and blockage dynamics) to optimize the communication performance? How much do we gain by doing so?} To address these questions, we envision the use of adaptive communication strategies and their formulation via partially observable (PO) Markov decision processes (MDPs).

We consider two base stations
 (BSs) serving a mobile user (MU)  on both sides of a road link. At any time, the MU is associated with one of the two BSs (the serving BS). To enable directional data transmission (DT), the serving BS performs beam training (BT); to compensate for blockage, it performs handover (HO) to the other BS on the opposite side of the road link.
The goal is to design the BT/DT/HO strategy,  so as to optimally trade-off the throughput delivered to the MU and the average power consumption of BS.
We formulate the optimization problem as a POMDP,
and develop an approximate dynamic programming algorithm based on PERSEUS~\cite{DBLP:journals/corr/abs-1109-2145}.
 Our numerical evaluations based on a Gauss-Markov mobility model  demonstrate that the PERSEUS-based policy performs very closely to a genie-aided upper bound in which 
 the position of the MU and the blockage states are known,
and outperforms a baseline scheme with periodic beam training by up to 55\% in spectral efficiency.
 Motivated by the structure of the PERSEUS-based policy, we design an adaptive heuristic policy with low computational cost,
 and show numerically that  it incurs a small 10\% degradation in spectral efficiency compared to the PERSEUS-based policy.

{\bf Related Work}: 
In the past decade, the design of 
beam training schemes for mm-wave systems has been the focus of extensive research,
ranging from beam sweeping \cite{michelusi2018optimal}, estimation of angles of arrival (AoA) and of departure (AoD) \cite{marzi}, to \mbox{data-assisted} schemes \cite{inverse_finger}, and  feedback-based schemes \cite{TWC2019}.
 Despite their simplicity, the overhead incurred by these algorithms may ultimately offset the benefits of beamforming in highly mobile environments~\cite{choi2016millimeter}. 
 \emph{In this paper, we contend that leveraging  a priori information on the vehicle's mobility as well as blockage dynamics may greatly improve the performance in vehicular communications}~\cite{va2016beam}. 
To this end, in \cite{michelusi2018optimal}, we designed optimal beam-sweeping schemes based on a worst-case mobility pattern.
 In~\cite{scalabrin2018beam}, we designed adaptive strategies for BT/DT that leverage a priori mobility information, but with no consideration of blockage, hence no handover. In this work, we exploit both mobility and blockage dynamics to design adaptive communications schemes via POMDP.
Related work that applies learning techniques to \mbox{mm-wave} networks includes~\cite{va2018online,second-best, alkhateeb2018machine}, 
  revealing a growing interest in the design of adaptive communication policies that exploit side information to enhance the overall network performance. 
For instance, \emph{contextual information} is exploited in
\cite{va2018online}  to reduce the training overhead,
and the feedback is used in \cite{second-best} to improve the beam search in the next rounds.
However, these works neglect the impact of realistic mobility and blockage processes on the performance.
	In~\cite{alkhateeb2018machine}, the serving BS predicts blockages using past observations,
	and proactively performs handover to another BS with highly probable LOS link. 
However, the MU speed is randomly selected from a predefined set of values, and thus does not follow a realistic mobility process.
Compared to this line of works, in this paper, we design adaptive communication strategies 
that leverage statistical information on the mobility and blockage processes in the 
selection of BT/DT/HO actions, with the goal to optimize the average long-term communication performance of the system. This approach is in contrast to strategies that
either lack a mechanism to perform handover \cite{va2018online,second-best}, or assume a non realistic mobility pattern in their design \cite{alkhateeb2018machine}.

\vspace{-3mm}

\section{System Model}
\label{sec:System_Model}
\begin{figure}[t]	
	\centering
	\includegraphics[trim = 0 0 0 20,clip,width=.5\columnwidth]{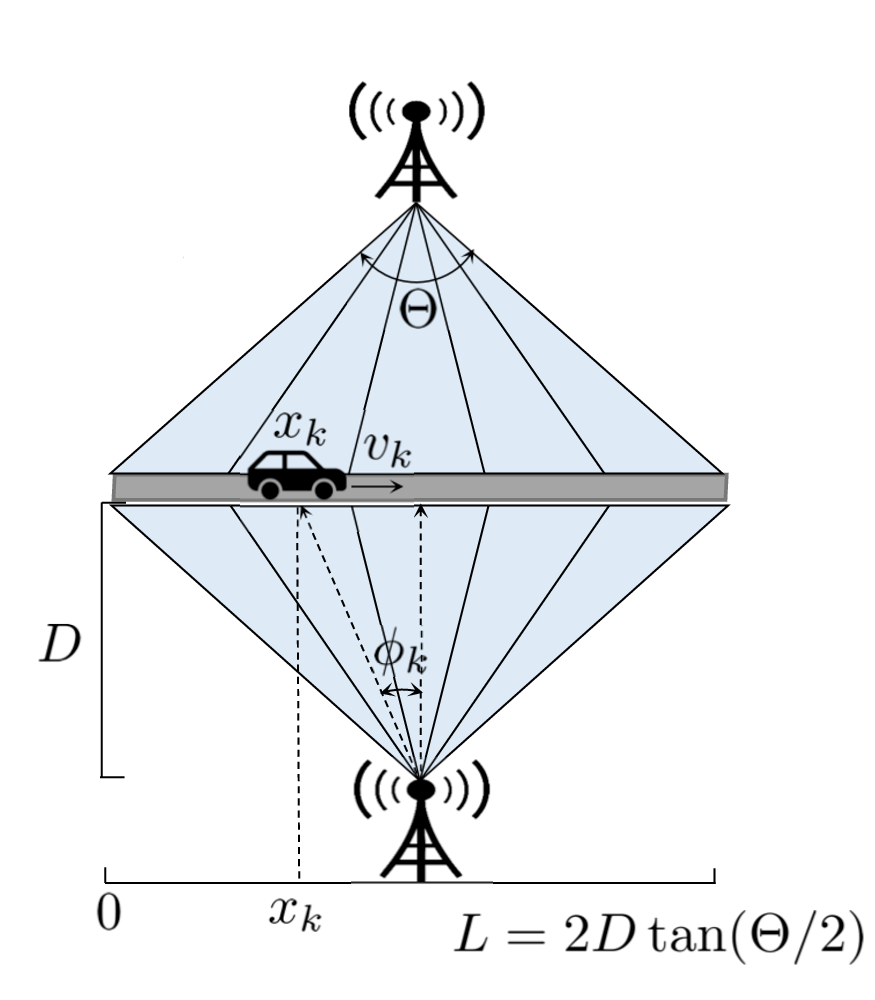}
	\caption{A cell deployment with BSs on both side of road.}\label{figure:Fig_scenario}
\end{figure}

We consider the scenario depicted in Fig.~\ref{figure:Fig_scenario}, where
 two BSs on both sides of a road link serve a MU moving along it. At any time, the MU is associated with one BS, denoted as the \emph{serving BS}, which performs data transmission (DT) to the MU using beamforming to create a directional link, along with beam training (BT) to maintain alignment.
The communication link between the serving BS and the MU is subject to time-varying blockage, which causes the signal quality to drop abruptly and DT to fail. 
To compensate for it, the serving BS may perform handover (HO) to the other BS on the opposite side of the road link, which then continues the process of BT and DT, until either another blockage event is detected, or the
 MU exits the coverage area of the two BSs.
 In this context, we investigate the design of the BT/DT/HO strategy,
so as to optimize a trade-off between maximizing the throughput delivered to the MU and minimizing the power consumption of the BS during a transmission episode, defined as the time interval between the two instants when the MU enters and exits the coverage area of the two BSs. 
Both BSs are at a distance $D$ from the road segment, symmetrically with respect to the road, and use a discrete set of narrow beams to communicate with the MU.
To this end, the road segment covered by the two BSs, of length $L{\triangleq} 2D\tan(\Theta/2)$ and angular range $\Theta$, is partitioned into $S$ sectors of equal length $\Delta_{\rm s} {=}L/S$, indexed by $s{\in}\mathcal S{\equiv}\{1,\dots,S\}$. Each sector is then associated with one transmission beamformer $\mathbf c^{(s)}$, 
with angular support
\begin{align}
\nonumber
\Phi_s{=}\bigg[ &\arctan\frac{(s{-}1)\Delta_{\rm s}-L/2}{D},\arctan\frac{s\Delta_{\rm s}-L/2}{D}\bigg],\ \forall s{\in}\mathcal S,
\end{align}
and beamwidth $\theta_s{=}|\Phi_s|$, so that the ensemble of all beams span the entire angular region covered by the two BSs. $\mathbf c^{(s)}$ can be defined with a proper beam design, as done in the numerical results in Sec. \ref{sec:numres} with the algorithm of \cite{noh}.
\\
\indent Time is discretized into \mbox{time-slots} of duration $\Delta_{\rm t}$, 
corresponding to a beacon signal during BT or a data fragment during DT.
Let $Z_k {\in} \bar {\mathcal S} \triangleq\mathcal S {\cup} \{ \bar s\} $ denote the sector 
occupied by the MU at time $k$, where $Z_k{=}\bar s$ indicates that the MU exited the coverage area of the BSs.
As a result of mobility of the MU, we model $Z_k$ as a discrete-time Markov chain over $\bar {\mathcal S}$, with transition probabilities
$\mathbf  P_{ss'} = \mathbb P(Z_{k+1} = s'|Z_k=s)$.
In the numerical results, we 
estimate $\mathbf  P$ from time-series generated with the Gauss-Markov mobility model,
in which the position $x_k$ and speed $v_k$ of the MU evolve as
\begin{align}
\label{GMmodel}
&v_k=\gamma v_{k-1}+(1-\gamma)\mu_v + \sigma_v\sqrt{1-\gamma^2} \tilde{v}_{k},\\
&x_k=x_{k-1}+\Delta_{\rm t}v_{k-1},
\end{align}
where $\mu_v$ and $\sigma_v$ are the average and standard deviation of $v_k$; $\gamma$ is a memory parameter and $\tilde{v}_{k} {\sim} \mathcal{CN}(0,1)$, i.i.d. over $k$.
\indent Within the $k$th \mbox{time-slot} of duration $\Delta_{\rm t}$, $L$ symbols each of duration $\Delta_{\rm t}/L$ are transmitted by the serving BS, denoted by the index $I_k\in\{1,2\}$.
 Let $\mathbf x_k\in\mathbb C^L$ be the signal transmitted such that $\mathbb E[\Vert \mathbf x_k\Vert_2^2]=L$. 
 Assuming isotropic reception at the MU, the received signal is expressed as
\begin{align}
\label{eq:signal_model}
\mathbf y_k = \sqrt{P_k} \mathbf{h}_k \mathbf{c}_{k}\mathbf x_k +\mathbf w_k,
\end{align}
where $P_k$ is the transmit power of the serving BS; $\mathbf{h}_k{\in} \mathbb{C}^{1\times M_{\rm tx}}$ is the channel vector;
$M_{\rm tx}$ is the number of antenna elements at each BS;
 $\mathbf{c}_{k} {\in} \mathbb{C}^{M_{\rm tx}\times 1}$ with $\Vert\mathbf{c}_{k}\Vert_2^2=1$
is the beamforming vector; $\mathbf w_k {\sim} \mathcal{CN}(0, \sigma_w^2\mathbf I)$ with \mbox{$\sigma_w^2=N_0W_{\rm tot}$} is
additive white Gaussian noise (AWGN), $N_0$ is the noise power spectral density, $W_{\rm tot}$ is the signal bandwidth.
\\
\indent In this paper, we model the channel as a single LOS path with binary blockage 
 state $b_k^{(i)}{\in}\{0,1\}$ \cite{blockage},
\begin{align}
\label{channel}
\mathbf{h}_k=\sqrt{M_{\rm tx}}b_k^{(I_k)} h_k \mathbf{d}_{\rm tx}(\psi_k)^H,
\end{align} 
where $b_k^{(i)}{=}1$ if the LOS path of BS $i$ is unobstructed, $b_k^{(i)}{=}0$ otherwise;
  $\mathbf{d}_{\rm tx}(\psi_k) {\in} \mathbb{C}^{M_{\rm tx}}$ is the BS array response vector with $\Vert\mathbf{d}_{\rm tx}(\psi_k)\Vert_2{=}1$; $\psi_k{\triangleq} \sin(\phi_k){=(x_k{-}L/2)/d_k}$ is the spatial angle corresponding to the AoD (computed with respect to the perpendicular to the array) $\phi_k {\in} [-\Theta/2,\Theta/2]$ in slot $k$; the term $h_k {\sim} \mathcal{CN}(0,\sigma_h^2)$ is the complex channel gain of the LOS component, i.i.d. over slots,
 with $\sigma_h^2{=}1/\ell(d_k)$; $\ell(d_k){=}[{4\pi d_k}]^2/\lambda_c^2$ denotes the distance-dependent path loss,
 as a function of the MU-BS distance
$d_k {=} d(\phi_k){=}D \sqrt{1{+}\tan(\phi_k)^2}$ (see Fig. \ref{figure:Fig_scenario});
$\lambda_c{=}c/f_c$ is the wavelength at carrier frequency $f_c$.
\\
\indent Letting $G_{\rm tx}(\mathbf{c},\psi)=M_{\rm tx}|\mathbf{d}_{\rm tx}(\psi)^H\mathbf{c}|^2$
be the beamforming gain of the serving BS and $\Theta_{\rm tx}=\angle{\mathbf{d}_{\rm tx}(\psi)^H\mathbf{c}}$ be its phase,
the signal received at the MU in slot $k$ can be expressed as 
\begin{align}
\label{sigmodel}
\mathbf y_k = \sqrt{P_k} b_k^{(I_k)}h_k \sqrt{G_{\rm tx}(\mathbf{c}_{k},\psi_k)} e^{j\Theta_{\rm tx}}\mathbf x_k + \mathbf w_k.
\end{align}
We use the sectored-antenna model, i.e., $G(\mathbf c^{(s)},\psi_k)/ d(\phi_k)^2$ is constant 
within the main-lobe
 $\phi_k{\in}\Phi_s$, so that, letting \mbox{$\Gamma\triangleq \frac{\lambda_c^2}{8\pi \sigma_w^2\Delta_s D}$},
  the average SNR when $\phi_k {\in} \Phi_s, b_k^{(I_k)}{=}1$ (alignment and no-blockage) can be shown to be
\begin{align}
\label{eq:P_t_2}
{\rm SNR}_k = \Gamma P_k,
\end{align}
This result is in line with the intuition that larger distances are achievable via smaller beamwidths, as also observed in~\cite{7744807}.
 If $\phi_k {\not\in} \Phi_s$ or $b_k^{(I_k)}{=}0$ (mis-alignment or blockage),
${\rm SNR}_k = \rho \Gamma P_k,
$
where $\rho {\in} (0,1)$ is the side- to main-lobe gain ratio, which is numerically found from the gain pattern. 
  \\
\indent Finally, the blockage state $b_k^{(i)}$ is modeled as a Markov chain with transition probabilities
\begin{align}
\label{markov}
&\mathbf P_{b\to b'}^{(i)}\triangleq\mathbb P(b_{k+1}^{(i)}=b'|b_k^{(i)}=b),\ \forall b,b'\in\{0,1\}.
\end{align}
The processes $\{b_k^{({i})},k\geq0\},i\in\{1,2\}$ evolve independently of each other, with Markov dynamics \eqref{markov}. The independence assumption is motivated by the fact that the two BSs are on opposite sides of the road segment, hence they experience different types of obstructions between the MU and the BS.
We now introduce the BT and DT operations.\\
\textbf{BT phase:} At the start of a BT phase, the BS selects a set of sectors $\hat {\mathcal S}_{\rm BT}$ over which it will send the beacons $\mathbf x_k$ for BT,
and a target SNR, ${\rm SNR}_{\rm BT}$.
The beacon transmission is done sequentially, using one slot for each sector in the set $\hat {\mathcal S}_{\rm BT}$. Therefore, the duration of the BT phase is $T_{\rm BT}\triangleq|\hat {\mathcal S}_{\rm BT}|+1$, which includes the last slot for feedback signaling from the MU to the BS.
 Let $i{\in}\{0,\ldots,T_{\rm BT}-2\}$ be the $i$th timeslot during the BT phase, and $\hat s_i\in\hat {\mathcal S}_{\rm BT}$ be the sector covered by the BS. At the MU, the received signal $\mathbf{y}_{k+i}$ is processed using a matched filter to generate the output
  \vspace{-1mm}
 \begin{align}
 z_{\hat s_i}=\zeta(\mathbf{x}_{k+i},\mathbf{y}_{k+i})\triangleq\frac{|\mathbf{x}_{k+i}^H \mathbf{y}_{k+i}|^2}{N_0 W_{\rm tot} \Vert\mathbf{x}_{k+i}\Vert_2^2}.
 \label{zsi}
\end{align}
 Upon collecting the sequence $\{z_{\hat s},\forall \hat s\in\hat {\mathcal S}_{\rm BT}\}$,
the MU generates the feedback signal as
\begin{align}
\label{btfb}
\!\!\!Y_k = \begin{cases}
{\hat s}^*\triangleq \arg\max_{{\hat s}\in\hat{\mathcal S}_{\rm BT}}z_{\hat s},& \max_{{\hat s}\in\hat{\mathcal S}_{\rm BT}} {z_{\hat s}}>\eta_{BT},\\
\emptyset,& \max_{{\hat s}\in\hat{\mathcal S}_{\rm BT}} {z_{\hat s}}\leq\eta_{BT}.
\end{cases}
\end{align}
 In other words, if all the matched filter outputs are below a threshold $\eta_{BT}$, the feedback $\emptyset$ is reported, indicating that no beam is deemed sufficient to carry data transmission, either due to blockage ($b_{k}^{(I)}=0$), or mis-alignment ($Z_{k}\not\in\hat{\mathcal S}_{\rm BT}$).
Otherwise, the ID of the strongest beam $\hat s^*$ is reported. 
\\
\textbf{DT phase:} At the start of the DT phase, the BS selects a sector $\hat s\in \mathcal S$ over which it performs DT for $T_{\rm DT} -1$ slots, along with a target average SNR at the receiver ${\rm SNR}_{\rm DT}$ and a target transmission  rate $\bar R_{\rm DT}$; an additional slot is used for the feedback signal from the MU to the BS, as described below, so that the overall duration of the DT phase is $T_{\rm DT}$. We assume that a fixed fraction $\kappa\in(0,1)$ out of $L$ symbols in each slot is used for channel estimation. 
Then, under alignment ($s=\hat s$ and $b_I=1$), 
and assuming that channel estimation errors are negligible compared to the noise level (which can be achieved with a sufficiently long pilot sequence $\kappa L$),
from the signal model \eqref{sigmodel}, we find that outage occurs if
\begin{align} 
W_{\rm tot} \log_2 (1+|h_k|^2\ell(d_k){\rm SNR}_{\rm DT})<\bar R_{\rm DT},
\end{align}
(note that $\mathbb E[|h_k|^2\ell(d_k)]=1$) yielding the outage probability
\begin{align*} 
\mathbb P_{\text{OUT}}(\bar R_{\rm DT},{\rm SNR}_{\rm DT})
=
1-\exp\Big\{-{\rm SNR}_{\rm DT}^{-1}(2^{\frac{\bar R_{\rm DT}}{W_{\rm tot} }}-1)\Big\}.
\end{align*}
 In this paper, we design $\bar R_{\rm DT}$ based on the notion of $\epsilon-$outage capacity, i.e., $\bar R_{\rm DT}$ is the largest rate such that $\mathbb P_{\text{OUT}}(\bar R_{\rm DT},{\rm SNR}_{\rm DT}) \le \epsilon$, for a target outage probability $\epsilon<1$. Setting $\mathbb P_{\text{OUT}}=\epsilon$, this can be expressed as 
 \begin{align} \label{eq:R_t}
\bar R_{\rm DT}{=}C_{\epsilon}({\rm SNR}_{\rm DT}){=}W_{\rm tot} \log_2\left(1{-}{\rm SNR}_{\rm DT} \ln(1-\epsilon)\right),
\end{align}
so that the average throughput is
\begin{align}
\label{thr}
\mathcal T(\epsilon,{\rm SNR}_{\rm DT})\triangleq(1-\kappa)(1-\epsilon)C_{\epsilon}({\rm SNR}_{\rm DT}),
\end{align}
 where $(1-\kappa)$ takes into account the overhead due to channel estimation. 
 Subsequently, we select $\epsilon$ to maximize $\mathcal T$,
 i.e., given ${\rm SNR}_{\rm DT}$, $\epsilon$ is chosen as the unique fixed point of 
 $\mathrm d \mathcal T(\epsilon,{\rm SNR}_{\rm DT})/\mathrm d\epsilon=0$.
 We denote the corresponding throughput maximized over $\epsilon$ as 
 $\mathcal T^*({\rm SNR}_{\rm DT})$.
\\
\indent We envision a mechanism in which the pilot signal transmitted in the second last slot of the DT phase (the most recent) is used to generate the binary feedback signal $Y{\in}\{\hat s,\emptyset\}$, transmitted by the MU back to the BS in the last slot of the DT phase. Similarly to the BT feedback, 
\begin{align}
\label{fbdt}
 Y_k = \begin{cases}
\hat s, &\zeta(\mathbf{x}_{k+T_{\rm DT} -2}^{(p)},\mathbf{y}_{k+T_{\rm DT} -2}^{(p)})>\eta_{DT}\\
\emptyset, &\zeta(\mathbf{x}_{k+T_{\rm DT} -2}^{(p)},\mathbf{y}_{k+T_{\rm DT} -2}^{(p)})\leq\eta_{DT},
  \end{cases}
  \end{align}
 based on the pilot signal $\mathbf{x}_{k+T_{\rm DT} -2}^{(p)}$ (of duration $\kappa L$)
  and on the corresponding signal $\mathbf{y}_{k+T_{\rm DT} -2}^{(p)}$ received on the second last slot of the DT phase, so that
  $Y{=}\hat s$ denotes beam-alignment, whereas $Y{=}\emptyset$ denotes loss of alignment due to either mobility of the MU or blockage.
  For both BT and DT, the feedback distribution  is computed  in closed-form in \cite{TVT2020}.
\vspace{-3mm}
\section{POMDP Formulation}
\label{sec:POMDP}
We now formulate the problem of jointly optimizing the BT, DT and HO strategy as a POMDP,
defined next.
 

\noindent \underline{\textbf{States:}} the state is denoted as ${u_k\triangleq}(Z_k,I_k,b_k^{(1)},b_k^{(2)})\in\mathcal U$ taking values from the set $\mathcal U=(\mathcal S\times\{1,2\}\times\{0,1\}^2)$, where
${Z_k}\in\mathcal S$ is the sector occupied by the MU, $I_{{k}}\in\{1,2\}$ is the index of the serving BS, and
$b_k^{(i)}\in\{0,1\}$ for $i\in\{1,2\}$ is the blockage state. We add the absorbing state $\bar s$ to denote the fact that the MU exited the coverage area of the two BSs, so that the overall state {space} is
$\mathcal{\bar{U}}=\mathcal U\cup\{\bar{s}\}.$

\noindent \underline{\textbf{Actions:}} the serving BS can perform
either BT, DT or HO actions. However, differently from standard POMDPs in which each action takes one slot, in this paper we generalize the model to actions taking multiple slots, as explained next.
\par Under action HO, the other BS becomes the serving one for the successive \mbox{time-slots}, until HO is chosen again. Its duration is denoted as $T_{\rm HO}$, modeling the delay to coordinate the transfer of the data traffic between the two BSs.
\par Under action BT, the serving BS chooses the set $\hat{\mathcal S}_{\mathrm{BT}}$ of sectors
 to scan and the target SNR {$\mathrm{SNR}_{\mathrm{BT}}$}. 
The duration of the BT action is $T_{\rm BT}{=}|\hat{\mathcal S}_{\mathrm{BT}}|{+}1$: $|\hat{\mathcal S}_{\mathrm{BT}}|$ slots for 
scanning the set of sectors $\hat{\mathcal S}$, and one slot for the  feedback from the MU to the serving BS.

Under action DT, the serving BS
selects the sector $\hat{s}$ covered,
the duration $T_{\rm DT}\geq 2$, and the target SNR $\mathrm{SNR}_{\mathrm{DT}}$ of the data communication session.
The transmission power is then determined via \eqref{eq:P_t_2}, and the transmission rate is given by \eqref{eq:R_t} to achieve $\epsilon$-outage capacity, so that the resulting expected throughput (in case of LOS and correct alignment) is $\mathcal T^*({\rm SNR}_{\rm DT})$.
The duration of the data communication session $T_{\rm DT}$ includes the second last slot 
to generate the feedback signal,
 which is fed back to the BS in the last slot.
 
 We denote the action as the 4-tuple $a {=} (c,\hat {\mathcal S}_c, {\rm SNR}_c,T_c)$, where $c{\in}\{{\rm HO},{\rm BT},{\rm DT}\}$ is the action class. For HO, we set $\hat{\mathcal S}_{\rm HO}{=}\emptyset$ and ${\rm SNR}_{\rm HO} {=} 0$. We denote the  action space  as $\mathcal A$.

	\noindent \underline{\textbf{Observations:}}
		upon selecting action $A_k{\in}\mathcal A$ of duration $T$ in slot $k$ and executing it in state $u_k{\in}\mathcal U$, the BS observes $Y_{k}$ from the set $\mathcal Y={\mathcal S}{\cup}\{\emptyset\}{\cup}\{\bar s\}$. The observation signal $Y_{k}{=}\bar s$ denotes that the MU exited the coverage area of the two BSs, hence the episode terminates; otherwise, $Y_{k}$ denotes the feedback signal after the action is completed, as described earlier for the BT and DT actions in \eqref{btfb} and \eqref{fbdt} (we set ${Y}_{k}{=}\emptyset$ under the HO action).

	\noindent \underline{\textbf{Transition, Observation probabilities:}}
	\label{sec:state_obs_prob}
	 Let $\mathbb P(u',y|u,a) \triangleq\mathbb P({U}_{k+T}{=}{u}',Y_k{=}y|{U}_k{=}{u},A_k{=}a)$ be the probability of 
moving from {state} $u{\in}{\mathcal U}$ to {state} $u'{\in}\mathcal{\bar{U}}$
and observing $y{\in}\mathcal Y$ 
under action $a{\in} \mathcal{A}$ of duration $T$. Note that these probabilities are 
a function of the duration $T$ of the selected action $a$, and can 
be computed in closed-form based on the feedback distribution and state transition probabilities (see \cite{TVT2020}).

\noindent \underline{\textbf{Costs and Rewards:}}
for every state action pair $(u,a)$, we let
$r(u,a)$ and $e(u,a)$ be the expected number of bits transmitted from the BS to the MU and the expected energy  cost, respectively.
Under the HO and BT actions, we have that $r(u,a)=0$ (since no bits are transmitted under these actions).
On the other hand, under the DT action
$a{=}({\rm DT},\{\hat s\}, \mathrm{SNR}, T_{\rm DT})$
 (of duration  $T_{\rm DT}$, SNR ${\rm SNR}$, over sector $\hat s$), the expected throughput in
  the $t$th communication slot is $\mathcal T^*({\rm SNR})$,
  provided that there is correct alignment and no blockage ($Z_{k+t}{=}\hat s$ and 
$b_{k+t}^{(I)}{=}1$); otherwise, outage occurs and the expected throughput is zero. Hence,
the total expected traffic delivered
over the entire communication session is
\begin{align}
\label{eq:reward}
&r((s,I,b_1,b_2),(\mathrm{DT},\{\hat s\}, \mathrm{SNR}, T_{\rm DT}))
\\ 
&{=}
\mathcal T^*({\rm SNR})
\sum_{t=0}^{T_{\rm DT}-2}\mathbb P(Z_{k+t} {=} \hat s,b_{k+t}^{(I)}{=}1|Z_k{=}s,b_{k}^{(I)}{=}b_I).
\nonumber
\end{align}
The energy cost under action $a$ with SNR ${\rm SNR}$
 is expressed from \eqref{eq:P_t_2} as
 (note that ${\rm SNR}{=}0$ and $e(u,a){=}0$ under HO)
\begin{align}
e(u,a)
=
\frac{\Delta_{\rm t}}{\Gamma}{\rm SNR}(T-1).
\label{eq:cost}
\end{align}
Note that the last slot is reserved to the feedback transmission, which incurs no energy cost for the BS.
We opt for a Lagrangian formulation to trade-off 
cost $e(u,a)$ and reward $r(u,a)$,
and we define $\mathcal{L}(u,a){=}r(u,a){-}\lambda e(u,a)$ for
$\lambda{\geq} 0$.


\noindent \underline{\textbf{Policy and Belief updates:}}
Since the agent cannot directly observe the system state $u_k$, we introduce the notion of \emph{belief}
 $\beta \in \mathcal{B}$, i.e., the probability distribution over system states, given the information collected so far at the BS.
Given $\beta$, the serving BS selects an action $a$ according to a policy $a=\pi(\beta)$, part of our design in Sec. \ref{sec:Problem_Optimization};
then, upon executing the action $a$ and receiving the feedback signal $y$, the BS updates the belief for the next decision interval according to Bayes' rule as
\begin{equation*}
\beta'(u')  = \mathbb P(u'\mid y, a, \beta)  = \frac{\sum_{u \in \mathcal{U}} \mathbb P(u',y|u,a)\beta(u)}{
\sum_{u\in \mathcal{U}} {\sum_{u''\in \bar{\mathcal{U}}}\mathbb P(u'',y|u,a)}\beta(u)
},
\end{equation*}
where $\mathbb P(u',y|u,a)$ is the conditional joint state transition and observation probability \cite{TVT2020}.
\vspace{-3mm}
\section{Optimization Problem}
\label{sec:Problem_Optimization}


Our goal is to determine a policy $\pi$ (i.e., a map from beliefs to actions) that maximizes
a trade-off between throughput and average power, 
${\bar{V}^{\pi}{\triangleq}}\bar T^{\pi}{-}\lambda\bar P^{\pi}$, starting from a given initial belief $\beta_0=\beta_0^*$ at time $0$.
Using Little's Theorem \cite{LittlesTheorem}, these metrics can be expressed as
\begin{align}
\label{metrics}
\bar T^{\pi}\triangleq \frac{\bar R_{\rm{tot}}^{\pi}(\beta_0^*)}{\bar D_{\rm{tot}}(\beta_0^*)},\ 
\bar P^{\pi}\triangleq \frac{\bar E_{\rm{tot}}^{\pi}(\beta_0^*)}{\bar D_{\rm{tot}}(\beta_0^*)},
\ 
\bar{V}^{\pi}\triangleq \frac{\bar V_{{\rm{tot}}}^{\pi}(\beta_0^*)}{\bar D_{\rm{tot}}(\beta_0^*)},
\end{align}
where $\bar D_{\rm{tot}}(\beta_0^*)$ is the expected episode duration, function of the mobility process but independent of policy $\pi$,
$$
[\bar R_{\rm{tot}}^{\pi}(\beta),\bar E_{\rm{tot}}^{\pi}(\beta)]{\triangleq}
\mathbb{E}_{\pi} \Big[ \sum_{t=0}^\infty [r({u_t,a_t}),e({u_t,a_t})] \Big| \beta_0{=}{\beta}\Big]
$$
are the total 
expected number of bits transmitted 
and the  total expected energy cost  during an episode,\footnote{{Note that the convergence of these series  is guaranteed by the presence of the absorbing state $\bar s$, i.e., the MU exits the coverage area at some point.}}
and
$$
\bar V_{{\rm{tot}}}^{\pi}(\beta){=}\bar R_{\rm{tot}}^{\pi}(\beta){-}\lambda\bar E_{\rm{tot}}^{\pi}(\beta)
=\mathbb{E}_{\pi}\!\bigg [ \!\sum_{t=0}^\infty\!\!\mathcal{L}(u_t,a_t)\Big| \beta_0 {=}\beta\!\bigg ]
.$$
Then, the optimization problem
starting from the initial belief $\beta_0=\beta_0^*$ is expressed as
$$
\!\max_{\pi}\ 
\bar {V}^{\pi}(\beta_0^*)
{=}
\frac{1}{\bar D_{\rm{tot}}(\beta_0^*)}\max_\pi\bar V_{{\rm{tot}}}^{\pi}(\beta_0^*) .
$$
It is well known that the optimal value function uniquely satisfies Bellman's optimality equation~\cite{DBLP:journals/corr/abs-1109-2145}
$V^*{=}H[V^*]$, where

\noindent we have defined the operator $\hat V{=}H[V]$ as
$$
\hat V(\beta){=}\max_{a \in \mathcal{A}} 
 \sum_{u \in \mathcal{{U}}}\!\beta(u)\!
 \bigg [\! \mathcal{L}(u,a){+}\!\sum_{y,u'}
 \mathbb P(u',y|u,a) V({\mathbb B(y,a,\beta)})\!\bigg ]\!,
$$
$\forall \beta\in\mathcal B$,
and the maximizer is the optimal policy $\pi^*(\beta)$.
 The optimal value function $V^*$ can be arbitrarily well approximated via the value iteration algorithm $V_{n+1}{=}H[V_n]$, where $V_0(\beta){=}0,\forall \beta$. 
Moreover, $V_n$ is
 a piece-wise linear and concave function~\cite{DBLP:journals/corr/abs-1109-2145}, so that
  it can be expressed by a finite set of hyperplanes $\mathcal Q_n\equiv\{\alpha_{n,i}\}_{i=1}^{A_n}$ of cardinality $A_n$, such that 
\begin{align}
\label{vn}
V_n(\beta)=\max_{\alpha\in\mathcal Q_n} \beta \cdot \alpha,
\end{align}
 where $\beta{\cdot}\alpha{=}\sum_u \beta(u)\alpha(u)$ denotes inner product. Each hyperplane $\alpha{\in}\mathcal Q_n$ is associated with an action $a_{\alpha}{\in} \mathcal{A}$,
 so that the maximizing hyperplane $\alpha^*$ in \eqref{vn} defines the policy $\pi_n(\beta){=}a_{\alpha^*}$.
It has been shown that $\mathcal Q_{n}$ 
grows 
doubly exponentially with the number of iterations,
 $A_{n+1}{=}|\mathcal Q_{n+1}|{=}
 |\mathcal A|^{|\mathcal Y|^n}$ \cite{Pineau2006PointbasedAF}.
For this reason, computing optimal policies for POMDPs is an intractable problem for any reasonably sized task.
 This calls for approximate solution techniques, e.g., PERSEUS~\cite{DBLP:journals/corr/abs-1109-2145}, which we introduce next.
\vspace{-5mm}
\subsection{Point-based Value Iteration (PBVI) for POMDPs}\label{sec:PERSEUS}

PERSEUS \cite{DBLP:journals/corr/abs-1109-2145} is an approximate PBVI algorithm for POMDPs. 
 The key idea is to define an approximate backup operator $\tilde{H}[\cdot]$ (in place of $H[\cdot]$), restricted to a discrete subset of belief points in $\tilde{\mathcal{B}}$, chosen as representative of the entire belief space $\mathcal B$;
 in other words, for a given value function $\tilde V_n$ at stage $n$, PERSEUS builds a value function $\tilde V_{n+1} = \tilde{H}[\tilde V_n]$ that improves the value of all belief points $\beta \in \tilde{\mathcal{B}}$,
without regard for the belief points outside of this discrete set, $\beta\notin\tilde{\mathcal{B}}$.
The goal of the algorithm is to provide a $|\tilde{\mathcal{B}}|$-dimensional set of hyperplanes $\alpha\in\mathcal Q$ and associated actions $a_{\alpha}$.
Given such set,
the value function at any other belief point $\beta\in\mathcal B$ is then approximated
via \eqref{vn}
as $\tilde V(\beta)=\beta\cdot\alpha^*$,  where $\alpha^*=\arg\max_{\alpha\in\mathcal Q} \beta\cdot\alpha$,
which defines an approximately optimal policy $\pi(\beta)=a_{\alpha^*}$.

\begin{figure}[!h]
	\centering
	\includegraphics[trim=20 0 20 10, clip,width=0.8\columnwidth]{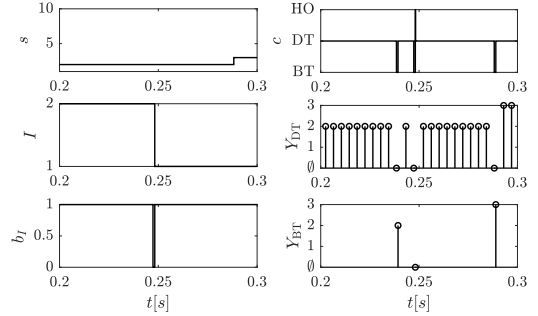}
	\caption{Execution of policy $\pi^*$.} 
	\label{figure:episode}
\end{figure}

The approximate backup operation of PERSEUS is given by Algorithm \ref{alg:alg_1},
which takes as input a set of hyperplanes $\mathcal Q_n$ and the corresponding actions, and outputs a new set $\mathcal Q_{n+1}$ along with their corresponding actions.
To do so:
in line 3, a belief point is chosen randomly from $\tilde{\mathcal{B}}_{\rm temp}$;
in lines 4-5, the hyperplane associated with each action $a\in \mathcal A$ is computed;
in particular, line 4 computes the hyperplane associated with the future value function $V_n(\mathbb B(y,a,\beta))$, for each possible observation $y$ resulting in the belief update $\mathbb B(y,a,\beta)$;
 line 5 instead performs  the backup operation to determine the new {one-step lookahead} hyperplane associated with each action;
 line 6 determines the optimal action that maximizes the value function for the current belief, yielding overall the value iteration update $V_{n+1}(\beta)=\max_a\mathbb E_{U,Y|a,\beta}[\mathcal L(U,a) +V_n(\mathbb B(Y,a,\beta))]$;
in lines 7-10, the new hyperplane and the associated action is added to the set $\mathcal Q_{n+1}$, but only if it yields an improvement in the value function $V_{n+1}(\beta)>\tilde V_{n}(\beta)$; otherwise, the previous hyperplane is used;
finally, lines 11-12 update the set  of un-improved beliefs based on the newly added hyperplane; only the belief points that have not been improved are part of the next iterations of the algorithm.
Overall, the algorithm guarantees monotonic improvements of the value function in the set 
$\tilde{\mathcal{B}}$,
 and continues until all beliefs have been improved and $\tilde{\mathcal{B}}_{\rm temp}$ is empty. Algorithm \ref{alg:alg_1} is then executed iteratively, until  convergence of the value function to a fixed point.
 \\
\indent To generate $\tilde{\mathcal{B}}$, we employ the \emph{Stochastic simulation and exploratory action} (SSEA) algorithm \cite{Pineau2006PointbasedAF}. After initializing $\mathcal{B}_0$, at iteration $n$, SSEA iteratively performs a one step forward simulation with each action in the action set, thus producing new beliefs $\{\beta_{a},\forall a\in\mathcal A\}$; hence, it computes the L1 distance between each new  belief point $\beta_a$ and its closest neighbor in $\mathcal{B}_n$, and adds the belief point $\beta_{a^*}$ farthest away from $\mathcal{B}_n$, so as to provide a wider coverage of the belief space. This expansion is performed multiple times to obtain $\tilde {\mathcal B}$.
\begin{algorithm}[t]
\DontPrintSemicolon
\SetNoFillComment
\SetKwFunction{Union}{Union}\SetKwFunction{FindCompress}{FindCompress}
\SetKwInOut{Input}{input}\SetKwInOut{Output}{output}
\caption{function {\rm PERSEUS}
\label{alg:alg_1}}
\Input {$\tilde{\mathcal{B}}$, $\mathcal Q_n$, $
\{a_\alpha^n,\alpha\in\mathcal Q_n\}$}
	\textbf{Init:} $\tilde V_{n+1}(\tilde\beta){=}-\infty,\forall \tilde\beta\in\tilde{\mathcal{B}}$; $\tilde{\mathcal{B}}_{\rm temp}\equiv\tilde{\mathcal{B}}$,  $\mathcal Q_{n+1}{=}\emptyset$; 
	 $\qquad\tilde V_{n}(\tilde\beta){\leftarrow}\max_{\alpha\in\mathcal Q_n} \tilde\beta {\cdot} \alpha$, maximizer $\alpha_{\tilde\beta},\ \forall \tilde\beta\in\tilde{\mathcal{B}}$;\;
	\While(\tcp*[f]{Unimproved beliefs}){$\tilde{\mathcal{B}}_{\rm temp} \neq \emptyset$}{
 Sample $\beta$ from $\tilde{\mathcal{B}}_{\rm temp}$;
		 For each action $a$, solve\\ $\alpha_{y,a}^*=\arg\max_{\alpha\in\mathcal Q_n}\mathbb B(y,a,\beta)\cdot\alpha,\ \forall y\in\mathcal Y$ and\\
		$\hat\alpha_{a}^*(u)=\mathcal L(u,a){+}\!\!
  \sum_{\hat u,y}\mathbb P(\hat u,y|u,a)
  \alpha_{y,a}^*(\hat u),\ \forall u$;\;
		 Solve $V_{n+1}(\beta)=\max_{a\in\mathcal A}
		\beta\cdot\hat\alpha_{a}^*$ and maximizing action $a^*$ and hyperplane $\hat\alpha=\hat\alpha_{a^*}^{*}$;\;
		\eIf(\tcp*[f]{$\hat\alpha$ improves value}){$V_{n+1}(\beta)>\tilde V_{n}(\beta)$}{
		$\mathcal Q_{n+1}\leftarrow\mathcal Q_{n+1}\cup\{\hat\alpha\}$; $a_{\hat\alpha}^{n+1}=a^*$ \tcp*[f]{add $\hat\alpha$ to $\mathcal Q_{n+1}$ and define action associated to $\hat\alpha$};\;
			}
		(\tcp*[f]{keep previous hyperplane $\alpha_\beta$}){
			 $\hat\alpha=\alpha_\beta$; $\mathcal Q_{n+1}\leftarrow\mathcal Q_{n+1}\cup\{\alpha_\beta\}$; $a_{\alpha_\beta}^{n+1}=a_{\alpha_\beta}^n$;\;
		}		
			  $\tilde V_{n+1}(\tilde \beta)\leftarrow\max\{\tilde \beta\cdot\hat\alpha, \tilde V_{n+1}(\tilde \beta)\},\forall \tilde \beta\in\tilde{\mathcal{B}}$;\;
			   \tcp*[l]{unimproved beliefs}
	  $\tilde{\mathcal{B}}_{\rm temp} {\leftarrow}
	\{\tilde \beta {\in} \tilde{\mathcal{B}}_{{\rm temp}} {:} \tilde V_{n+1}(\tilde \beta) {<} \tilde V_n(\tilde \beta) \}$;\;
	}	
  	\Return $\mathcal Q_{n+1}$, 
	$\{a_\alpha^{n+1},\forall\alpha\in\mathcal Q_{n+1}\}$\tcp*[f]{new hyperplanes and associated actions}\; 
	\end{algorithm}
\\
\indent After returning the set of hyperplanes $\mathcal Q_{n+1}$ and the associated actions $\{a_\alpha^{n+1},\forall\alpha\in\mathcal Q_{n+1}\}$, the (approximately) optimal action when operating under the belief $\beta$
can be computed as
 $$
 \pi^*(\beta)=a_{\alpha^*}^{n+1},\ \text{where }\alpha^*=
 	\arg\max\limits_{\alpha\in\mathcal Q_{n+1}}\beta\cdot \alpha.
 $$
\indent In Fig.~\ref{figure:episode}, we plot a time-series of
the evolution of state variables for 
a portion of an episode executed under the PERSEUS-based policy (Algorithm \ref{alg:alg_1}). The parameters used are listed in Table \ref{table1}.
Initially, the MU is known to be in sector $Z_0{=}1$, with LOS conditions for both BSs ($b_0^{(1)}{=}b_0^{(2)}{=}1$).
 We show a time-series for the sector index $Z_k$,
 index of the serving BS $I_k$, its blockage state $b_k^{(I_k)}$, the action class $c{\in}\{\rm{DT},\rm{BT},\rm{HO}\}$,
the BT feedback $Y_{\rm BT}$ as defined in \eqref{btfb}, and the DT feedback $Y_{\rm DT}$ as defined in \eqref{fbdt}. The action space for the DT time is set as $T_{\rm DT} {\in} \{ 10,20,40 \}$ and the power levels are set as  $P_{\rm BT}, P_{\rm DT}{\in}\{0,10,20,30,40\}$ (dBm). It can be observed in the figure that at 0.238s, 0.246s and 0.287s, NACKs are received after executing the DT action. After each one of these NACKs, the policy executes the BT action. If the BT feedback $ Y_{\rm BT} {\neq} \emptyset$, then DT is performed; otherwise, blockage is detected and  the  HO action is executed. 
    Next, we will present a heuristic policy that mimics this behavior.
\vspace{-7mm}
\subsection{Heuristic Policy}
\label{sec:heuristic} 
\begin{figure}[h]
	\centering
	\vspace{-4mm}
	\includegraphics[trim=0 15 0 10 , clip,width=0.7\columnwidth]{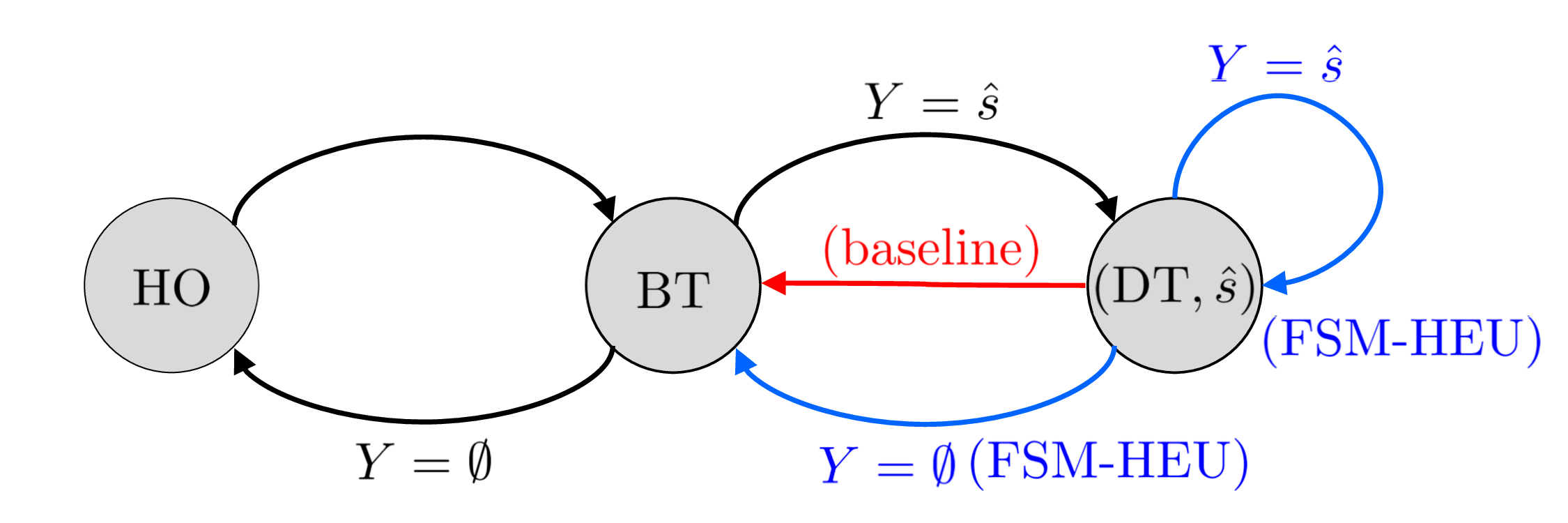}
	\caption{Finite state machine based on the observation signal $Y$. Black lines represent transitions under both FSM-HEU and baseline; blue and red lines represent transitions under FSM-HEU and baseline only, respectively.
		}\label{figure:Fig_ci}
\end{figure}

Note
 that Algorithm~\ref{alg:alg_1} incurs a huge computational cost especially for POMDP with large state and action spaces (hence large number of representative belief points). To remedy this, we propose a 
finite state machine based heuristic policy (FSM-HEU) that will be shown numerically to achieve \emph{near-optimal} performance.
 The key idea of FSM-HEU is that it selects the BT/DT/HO actions based solely on the last action executed and its observation signal, but not on the belief $\beta_k$. The behavior of this scheme can thus be described as a finite-state machine, depicted in Fig.~\ref{figure:Fig_ci} and described next. 

If the last action executed was a BT action, and the feedback signal is $Y=\hat s$ (see \eqref{btfb}),
then the BS  detects the strongest beam $\hat s$; 
hence the next action selected is DT over sector $\hat s$ (the strongest detected), of fixed duration $T_{\mathrm{DT}}$.
On the other hand, if the feedback signal is $Y=\emptyset$, the BS detects blockage and performs handover to the non-serving BS (action HO).

If the last action executed was DT on sector $\hat s$, and the feedback signal is ACK ($Y{=}\hat s$, see \eqref{fbdt}), then
 the BS infers that the signal is still sufficiently strong to continue DT on the same sector, and the same action is selected;
otherwise (NACK received, $Y{=}\emptyset$), 
 the BS detects a loss of alignment, hence the BT action with exhaustive search is selected.

Finally, if the last action executed was HO, then the new serving BS executes BT via exhaustive search to locate the MU. This procedure continues until the episode terminates.

To study its performance, note that the underlying system state $U_k$ and action $A_k$ form a Markov chain.
Letting $\mathbb P(a'|y,a)$ be the probability of generating the new action $a'$, given previous action $a$ and observation $y$, as given by the finite-state machine of Fig.~\ref{figure:Fig_ci}, the value fuction $V(u,a), \forall u,a$ is obtained by solving the following system of linear equations
$$
V(u,a){=}
 \mathcal{L}(u,a){+}\!\!\sum_{y{,}u'{,}a'}
   \!\! \mathbb P(u',y|u,a)
  \mathbb P(a'|y,a)
V(u',a'), \forall u,a.
$$
\begin{table}[t]
\footnotesize
\begin{center}
\begin{tabular}{|l|l|l|}
\hline
   Parameter & Symbol &  Value \\ \hline
  Number of BS antennas & $M_{\rm tx}$ & $128$ \\
  Angular BS coverage & $\Theta$& $90^\circ$ \\
  Slot duration & $\Delta_t$ & $100 \mu {\rm s}$ \\ 
  Distance of road to BS & $D$ & $20$m \\ 
  Bandwidth & $W_{\rm tot}$& $100$MHz \\
   Carrier frequency & $f_c$ & $30$GHz \\
    Noise psd &$N_0$ & $-163$dBm/Hz \\
    Fraction of DT slot for channel- & &\\
    estimation/hypothesis-testing & $\kappa$ & $0.01$  \\
    HO delay & $T_{\rm HO}$ & 1 slot \\
    LOS to blockage transition prob. & $\mathbf P_{1\to 0}$ & $1.25\times 10^{-4}$\\
    Blockage to LOS transition  prob.& $\mathbf P_{0\to 1}$ &$5\times 10^{-4}$ \\
    MU average speed & $\mu_v$ & $30$m/s\\
    MU speed standard deviation &$\sigma_v$ & 10\\
    MU mobility memory parameter & $\gamma$ & 0.2\\
    \hline
\end{tabular}
\normalsize
\caption{Simulation parameters.}
\label{table1}
\end{center}
\end{table}
 
\section{Numerical results}
\label{sec:numres}
In this section, we perform a numerical evaluation of the various algorithms proposed in this paper, with simulation parameters listed in Table \ref{table1}.The blockage transition probabilities given in the table correspond to steady state blockage probabilities $\pi_{B}^{(1)}{=}\pi_{B}^{(2)}{=}0.2$ and average blockage duration of 0.2ms. 
Using the throughput metric defined in \eqref{metrics}, the average spectral efficiency is 
 computed as $\bar T^{\pi}/W_{\rm tot}$[bps/Hz].
 We compare the performance of the proposed policies to a baseline scheme which performs periodic BT, unless blockage is detected (in which case it executes HO, see Fig. \ref{figure:Fig_ci}). 

\begin{figure}[t!]
	\centering
	\includegraphics[trim=10 0 10 10, clip,width=0.95\columnwidth]{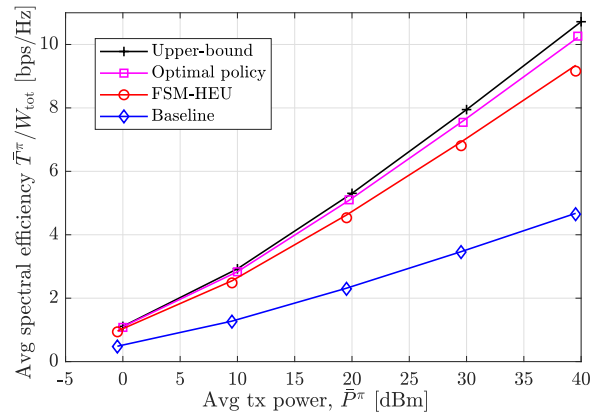}
	\caption{Average spectral efficiency versus average power consumption:
	analytical curves based on the sectored  antenna and mobility model (continuous lines) and simulation using analog beamforming and Gauss-Markov mobility (markers). 
	} 
		\label{figure:SEvP}
\end{figure}

 In Fig.~\ref{figure:SEvP}, we depict the average spectral efficiency against the average power consumption. For the FSM-HEU and baseline policies, we set $T_{\rm DT}{=}10$, and $P_{\rm BT}{=}P_{\rm DT}$ is varied from $0$dBm to $40$dBm.
  The upper-bound shown in the figure is obtained by
 a genie-aided policy that always executes DT with perfect knowledge of the state $(s,I,b_1,b_2)$ and hence its throughput performance can be upper bounded by
 $(1{-}\pi_{B}^{(1)}\pi_{B}^{(2)}){\mathcal T^*({\rm SNR}_{\rm DT})}$,
 i.e., it is $\mathcal T^*({\rm SNR}_{\rm DT})$ unless there is no LOS under both BSs (with steady-state probability $\pi_{B}^{(1)}\pi_{B}^{(2)}$)
 whereas its power consumption is given as $ (1-\pi_{B}^{(1)}\pi_{B}^{(2)}) P_{\rm DT}$. Note that this upper-bound is not attainable since it is found by assuming perfect knowledge of the state and ignoring the inefficiencies due to the time required to perform handover and transmit feedback. The PERSEUS-based policy $\pi^*$ yields the best performance with negligible performance gap with respect to the upper-bound. 
 It shows a performance gain of up to 11\% and 55\% compared to FSM-HEU and baseline, respectively. However, the baseline policy yields up to $50$\% degraded performance compared to FSM-HEU: in fact, the baseline scheme neglects the DT feedback and instead performs periodic BT, thus incurring significant overhead. We observe that the curves corresponding to analysis and the one based on simulation (based on the Gauss-Markov mobility model and beam design via \cite{noh}) closely match, thereby showing that the model introduced in the paper provides good abstraction of more realistic settings.
\vspace{-8mm}
\section{Conclusions}
\label{sec:Conclusions}
In this paper, we have investigated the design of beam-training/data-transmission/handover strategies for mm-wave vehicular networks. The mobility and blockage dynamics have been leveraged to obtain the approximately optimal policy via a POMDP formulation and its solution via a point-based value iteration (PBVI) algorithm based on PERSEUS~\cite{DBLP:journals/corr/abs-1109-2145}. Inspired by it, we have proposed a heuristic policy, which provides low computational alternatives to PBVI and exhibits performance comparable to the optimal policy obtained  via PBVI. Our numerical results demonstrate the importance of an adaptive design to tackle the highly dynamic environments caused by mobility and blockages in vehicular networks.
\bibliographystyle{IEEEtran}
\bibliography{IEEEabrv,bibliography} 

\end{document}